\documentstyle[aps,prl,epsf,multicol]{revtex}
\begin{document}
\draft
\title{Anomalous Roughness in Dimer Type Surface Growth}
\author{Jae Dong Noh$^{1,2}$, Hyunggyu Park$^{1,3}$,
        and Marcel den Nijs$^{1}$}
\address{$^1$ Department of Physics, University of Washington,
         Seattle, Washington 98195-1560, U.S.A.}
\address{$^2$ Center for Theoretical Physics, Seoul National University,
         Seoul 151-742, Korea}
\address{$^3$ Department of Physics, Inha University,
         Inchon 402-751, Korea}
\date{\today}
\maketitle

\begin{abstract}
We point out how geometric features affect the scaling properties
of non-equilibrium dynamic processes, by a model for
surface growth where particles can deposit and evaporate only in dimer form,
but dissociate on the surface.
Pinning valleys (hill tops) develop spontaneously
and the surface facets for all growth (evaporation) biases.
More intriguingly, the scaling properties of the rough one dimensional
equilibrium surface are anomalous.
Its width, $W\sim L^\alpha$, diverges with system size $L$,
as $\alpha=\frac{1}{3}$ instead of the conventional universal
value $\alpha=\frac{1}{2}$.
This originates from a topological non-local evenness constraint on the
surface configurations.

\end{abstract}


\begin{multicols}{2}
\narrowtext

The theory of non-equilibrium dynamic statistical processes has developed
rapidly in recent years.  Driven systems display intriguing scaling properties
and can undergo various types of dynamic phase transitions ~\cite{general}.
Kardar-Parisi-Zhang type surface growth is an example~\cite{KPZ}.
There, the properties of the depositing (evaporating) particles are not
specified,
but are implicitly presumed to be geometric featureless monomers.
In surface catalysis type processes the geometric shape of the molecules
matters.
The onset of the catalytic process is associated with a so-called absorbing
state
dynamic phase transition~\cite{DP,DI,DI2}.
Monomers give rise to directed percolation  and dimers to
directed Ising type transitions.
Subtle geometric features are known to be important in equilibrium crystal
surface
phase transitions as well.
The competition between surface roughening and surface reconstruction
depends on
topological details of the crystal symmetry. Those determine
whether a reconstructed rough phase can exist or not~\cite{MdN-King}.
Geometric features are also important in diffusing particle systems.
The shape of diffusing particles introduces conservations and leads to
anomalous decay of particle density autocorrelations~\cite{MenonBarmaDhar}.
Therefore, the natural question arises, whether and how
the shape of the deposited particles influences the growth and
equilibrium properties of crystal surfaces.

Consider a crystal built from atoms of type $X$.
Assume that deposition always takes place in dimer form, $X_2$,
aligned with the surface.
The dimer attaches to 2 horizontal nearest neighbour surface sites.
It looses its dimer character after becoming part of the crystal.
Evaporation can take place only in $X_2$ molecular form,
but a different partner is allowed.
In this letter we study the
one dimensional (1D) version of this process.
This can apply to step shapes during step-flow type growth on vicinal surfaces.
The adsorbed particles do not diffuse in this version of our model.
However, topological features that drive our results are preserved,
even when monomer diffusion is allowed but limited to terraces.
Jumps across steps are unlikely due to Schwoebel barriers ~\cite{schwoebel}.
So our main results do not alter in systems with diffusion.

We describe the 1D surface configurations in terms of integer height
variables
$h_i=0,\pm1,\pm2,\cdots$.
They are subject to the so-called restricted solid-on-solid (RSOS)
constraint, $h_i-h_{i+1}=0, \pm 1$, and
periodic boundary conditions, $h_{L+i}=h_i$.
The dynamic rule is as follows.
First, select at random a bond  $(i,i+1)$.
If the two sites are not at the same height,
no evaporation nor deposition takes place.
If the two sites are at the same height,
deposition of a dimer covering both sites is attempted with
probability $p$, or evaporation of a dimer with probability $q=1-p$.
Processes are rejected if they would result in a violation of the RSOS
constraint.

\begin{figure}
\centerline{\epsfxsize=8cm \epsfbox{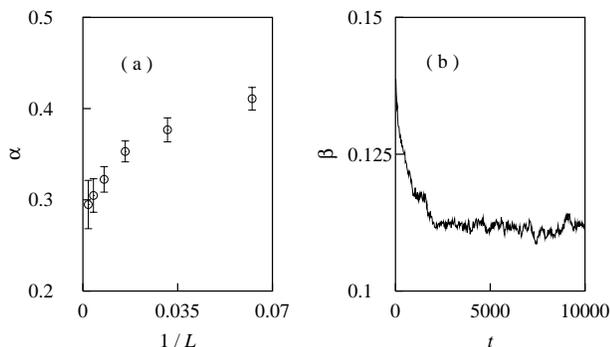}}
\caption{Finite size scaling estimates of the critical exponents
at the equilibrium point $p=q$ for:
(a) The stationary state surface width $W\sim L^\alpha$
using $L-2~L$ pairs for $L=2^4,\ldots,2^{9}$.
(b) The temporal surface width $W\sim t^\beta$,
using $t/10 - t$ pairs at  $L=2^{13}$.}
\label{width}
\end{figure}

The first surprise is that the surface always facets during growth and
evaporation,
although the surface is rough in equilibrium.
The second surprise is that the equilibrium surface width
$W\sim L^\alpha$ scales with an anomalous small exponent $\alpha= 0.29\pm
0.04$.
The data could be consistent with an even smaller value
due to the strong finite size scaling corrections in Fig.~\ref{width}(a).

1D surfaces, irrespective of being in equilibrium or in a
stationary growing (evaporating) state,
display, with only a few very specific exceptions,
the universal roughness exponent $\alpha=\frac{1}{2}$.
The up-down aspect of the steps become uncorrelated beyond a definite
correlation length,
and therefore the surface roughness obeys random walk statistics at large
length scales,
which implies  $\alpha=\frac{1}{2}$.
Our dissociating dimer deposition process circumvents this universal argument
by means of a novel type of non-local topological constraint.
The  dimer aspect implies that
all surface height levels must be occupied by an even number of particles.
However, due to the dissociative nature of the dimers
that information is not preserved locally.
The ``evenness" constraint is non-local.
At local length scales the surface looks the same as in monomer deposition
processes,
but the global surface is much less rough.
We checked this numerically.
Define a window of length $b$. The surface roughness scales
as $W\sim b^{\alpha}$, with $\alpha=\frac{1}{2}$ for $b\ll L$, but
crosses-over to
the global finite size scaling exponent $\alpha\simeq 0.29 \pm 0.04$ for
$b\to L$.

We performed a detailed numerical study of the properties of even-visiting
random walks~\cite{even-rw} that are
globally restricted to visit each site an even number of times.
The results, together with analytical scaling arguments, yield the value
$\alpha=\frac{1}{3}$.
This value lies within the numerical error bars in
Fig.~\ref{width}(a) for the  dimer deposition model.

The details of our random walk study are rather technical and
will be presented elsewhere~\cite{RRW}, but the essence
can be captured by the following intuitive scaling argument.
Consider the even-visiting random walks for time interval $0<t<T(=L)$.
We assign a defect variable to each site, to mark that it
has been visited by the random walker an odd/even number of times up to
time $t$. The even-visiting constraint is
satisfied when all defects disappear at time $T$.
Initially, the random walker does not feel the constraint 
and diffuses freely for $t < \tau_{free}\ll T$.
The defects are uniformly spread over a region of size 
$\xi\sim \tau_{free}^{1/2}$. Then it stops spreading and 
starts to heal the defects.
The healing time for a single defect in the region of size $\xi$
is order of $\xi^2$. By assuming that 
the healing process for each defect is independent, 
we estimate the total healing time $\tau_{heal} \sim \xi^{d+2}$
with spatial dimensionality $d$. As $\tau_{heal} \gg \tau_{free}$,
we conclude that $\xi\sim T^{1/(2+d)}$, i.e., $\alpha=1/3$ for $d=1$. 
Existence of a time scale $\tau_{free}$ explains the crossover behavior
of the surface roughness in the window length $b$.
Similarly, we can argue that the surface width
scales with $\alpha=\frac{1}{3}$ in generalized $n\geq 2$ dissociating
$n-mer$ type deposition processes.

The dynamic critical exponent $z$ at the equilibrium point follows from
how the surface width diverges as function of time, $W\sim t^\beta$.
We find numerically that $\beta =0.111\pm 0.002$, see Fig.~\ref{width}(b).
This suggests the value $z=3$, since $z=\alpha/\beta$.

\begin{figure}
\centerline{\epsfxsize=8cm \epsfbox{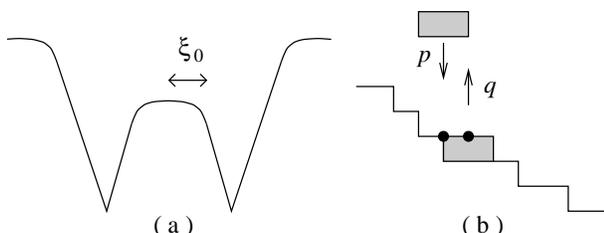}}
\caption{
Hill structures in the faceted growth phase.
(a) Schematic structure with two pinning valleys.
$\xi_0$ is the characteristic width of the hill tops.
(b) Active bonds (the filled circles)
along a local segment of a faceted ramp.
}
\label{Wconfig}
\end{figure}

The equilibrium surface roughness is unstable with respect to growth
and evaporation. It facets immediately.
This phase transition is second order.
The correlation lengths that characterize the faceted structure diverge.
Before addressing this issue we need to describe and explain the faceted phase.
The valleys in the growing surface are sharp and the hill tops rounded,
see Fig.~\ref{Wconfig}. This shape is inverted for $p<q$.
The faceting is caused by the spontaneous formation of pinning valleys
during growth (pinning hill tops during erosion, for $p<q$).
Consider for example, dimers on a flat surface for $p>q$.
Odd segments between them act as the nuclei of pinning valleys.
Such valleys can not be filled by direct deposition.
The only way to get rid of them is by lateral movement of the sub-hills.

In finite systems the surface grows in shocks.
An initial rough or flat configuration grows fast at first, but
pinning valleys appear randomly at all surface heights.
The interface develops into a rough faceted structure with many
sub-hills and growth almost stops.
From here on the surface advances only when sub-hills anneal out
by the lateral movement of the  pinning valleys.
The annealing time of a sub-hill scales exponentially with its size.
This exponentially slowing down healing process leads ultimately
to a faceted $W$ shape with only two remaining pinning valleys.
Their lifetime diverges exponentially with the lattice size. 
After their demise the surface experiences a growth spurt,
and the cycle restarts all over.

The mechanism for lateral movement of pinning valleys
is exchange of active bonds between ramps.
Active bonds are locations along the ramp where a dimer can deposit or
evaporate.
Most steps on the ramps are only one or two atomic units wide and
therefore dynamically inactive, see Fig.~\ref{Wconfig}.
Active bonds move up or down the ramps by deposition or evaporation of dimers.
The growth bias $p>q$ gives them an upward drift.
This must lead to an exponential distribution.

\begin{figure}
\centerline{\epsfxsize=8cm \epsfbox{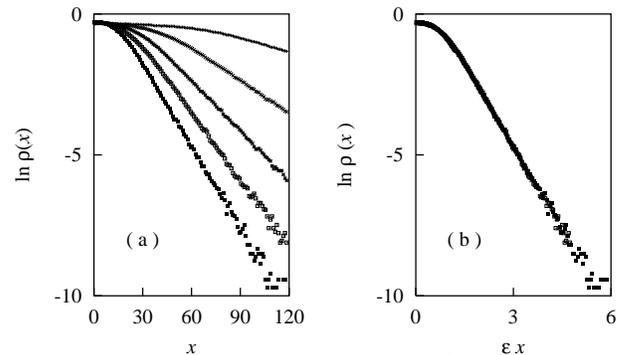}}
\caption{
Active bond distribution $\rho(x)$ at $L=257$ for
$p=0.505$, 0.51, 0.515, and 0.52 (from top to bottom).
Beyond $\xi_0$, $\rho(x)$  decays exponentially.
(b) Collapse of the data from (a) according to Eq.~(1), with $\beta_\rho=0$
and $\nu=1$,
demonstrating the hills preserve their shape.
}
\label{rho_x}
\end{figure}

Fig.~\ref{rho_x}(a)  shows the
logarithm of the active bond distribution, $ \rho(x)$,
versus the horizontal distance $x$ from the center of a hill top for
various values of $p>q$.
The straight lines for large $x$ confirm the exponential distribution of
active bonds along the ramps, $\rho(x)\sim \exp[-x/\xi_f]$.
We determined this from odd lattice sizes, in particular $L= 257$,
where the surface contains an odd number of pinning valleys
and therefore reaches a $V$ shaped stationary state in which it
remains pinned at all times.

Every now and then an active bond moves in the opposite direction,
against the flow, and reaches the valley bottom.
That pinning valley moves by two lateral lattice constants when the active bond
jumps across onto the other ramp.
The probability for this is very small, and scales exponentially
with the ramp length, but it is larger from the lower ramp, and
therefore the valley bottom moves in the direction of the lower hill, and
actually accelerates, because that lower hill keeps shrinking.

Near the rounded hill tops,
the surface remains highly active
and initially the active bond density does not decrease significantly with $x$.
This defines a second characteristic length scale, $\xi_0$,
representing the flatness of the rounded hill tops, see Fig.~\ref{Wconfig}.
Surprisingly, both lengths, $\xi_f$ and $\xi_0$, diverge at $p=q$.
Fig.~\ref{rho_x}(b) shows that the curves in Fig.~\ref{rho_x}(a)
collapse  according to a single length scaling form
\begin{equation}
\rho(\epsilon, x) = \epsilon^{\beta_\rho} {\cal F}(\epsilon^{\nu} x)
\end{equation}
with $\nu=1.0(1)$ and $\beta_\rho=0.0(1)$.
This means that on approach of the $p=q$ critical point the hills
maintain their shape in the sense that  $\xi_0$ and $\xi_f$ diverge
simultaneously
and with the same exponent  $\xi_0\sim \xi_f\sim (p-q)^{-1}$.

It is surprising that both length scales of the
faceted phase diverge  at the equilibrium point.
The structure of the rough phase would be much more complex when one
of them remained finite.
This actually happens in the following generalization of the model.

Recently, Alon {\it et.~al.}~\cite{Alon} added to the conventional monomer
type
RSOS model growth the constraint that evaporation from flat segments is
forbidden.
It remains unclear how this can be experimentally implemented,
but the interesting aspect of their model is the presence of a
roughening transition,
belonging to the directed percolation (DP) universality class~\cite{DP},
and unconventional roughness properties at this transition.
Our model becomes a directed Ising (DI) type~\cite{DI,DI2} generalization
of this
when we disable digging on flat surface segments.
Modify the evaporation probability  $q$ to $rq$  when
both neighbours are at the same height as the update pair  $(i,i+1)$,
i.e., $h_{i-1}=h_i=h_{i+1}=h_{i+2}$.
At $r=0$, the no-digging limit,
it becomes impossible to dig into the crystal layers beneath the
current lowest exposed level. That level itself becomes frozen as well
when it fills-up completely.

\begin{figure}
\centerline{\epsfxsize=8cm \epsfbox{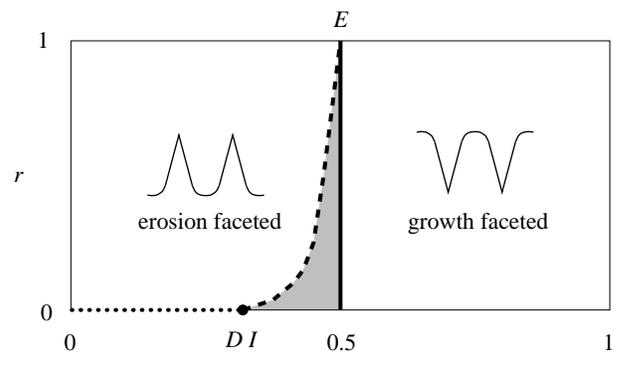}}
\caption{
Phase diagram of the generalized model:
The rough equilibrium point $E$ at $r=1$ broadens into a slowly growing
rough phase
(shaded) when the digging probability  $r$ is reduced.
The dotted line to the left of the DI critical point
at $r=0$ represents the smooth phase.
}
\label{phase_diagram}
\end{figure}

Figure~\ref{phase_diagram} shows the phase diagram.
The rough equilibrium point broadens into a rough phase (the shaded area).
Along the $DI-E$ phase boundary, the surface growth is zero.
Inside the rough phase the surface grows, but slowly.
Its scaling properties are complex
and obscured numerically by huge corrections to scaling.
The surface  width $W$ seems to grow logarithmically in time for $L>2^{10}$,
and maybe the stationary state width scales logarithmically as well,
but extracting the true scaling properties is a rather hopeless endeavor.

The origin of this complexity is easy to pin point.
The rough surface grows slowly,
although the bare coupling constants $p<q$ favours evaporation.
It performs a delicate balancing act.
The surface erosion picture from $r=1$ still holds along faceted ramp
segments.
There the surface evaporates due to the downward drift velocity of
active bonds along the slope, but this is frustrated by the emergence
of pinning hill tops.
The surface grows at flat surface segments due to an upward pressure
created by the reduced digging probability factor $r$,
but the formation of pinning valleys limits this.
Moreover, the non-local evenness constraint is at work as well.
Growth and evaporation are dynamically balanced
only along the faceting transition line $DI-E$ (Fig.~\ref{phase_diagram}).
Everywhere else the rough surface grows slowly.

The properties of the two faceted phases confirm the above intuitive picture.
The erosion faceted phase signals the local stability of eroding ramps
inside the rough phase, while the growth faceted phase indicates
that flat segments persist.
We have  numerical evidence showing that the active bond
characteristic length $\xi_f$ of the erosion faceted phase does
not diverge along the roughening transition line $DI-E$ for $r<1$.
This confirms that eroding ramps remain locally stable.
On the other side of the phase diagram,
the flatness length scale $\xi_0$ of the growth faceted phase
does not diverge along the $p=q$ roughening line.
This confirms the persistence of locally stable
flat segments inside the rough phase.

\begin{figure}
\centerline{\epsfxsize=8cm \epsfbox{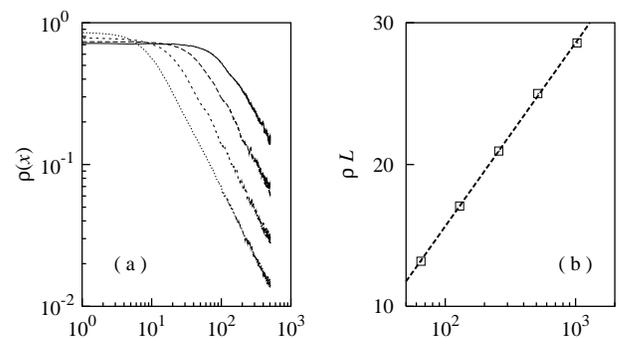}}
\caption{
Scaling at the $p=q$ transition line:
(a) Active bond distribution $\rho(x)$, along $p=q$
for system size $L=1025$
at $r=0.9, 0.8, 0.6$, and $0.2$ (ordered from the top down).
$\rho(x)\sim x^{-1}$ decays algebraically beyond
a non-diverging  hill top flatness length scale $\xi_0$.
(b) Scaling of the total active bond density
$\bar{\rho} \sim \ln L/L$ at $p=q$ and $r=0$.
}
\label{rho_xp=q}
\end{figure}

To illustrate this, we present some of the details of the latter.
Recall that along $r=1$, the active bond distribution $\rho(x)$
for different $p$ collapses onto a single curve (Fig.~\ref{rho_x}$(b)$).
For $r<1$ this fails.
At the transition point $p=q$, $\rho(x)$ scales
algebraically, as $\rho(x)\sim x^{-1}$, see Fig.~\ref{rho_xp=q}$(a)$,
but only beyond the central flat part of the hills.
The flatness length scale, $\xi_0$ remains finite.
Its value varies as $\xi_0\sim |1-r|^{-\nu}$, with $\nu= 1.0(1)$
along the line $p=q$.
In Fig.~\ref{rho_xp=q}$(a)$, $\xi_0$ can be as large as $\xi_0\simeq 40$.
This explains the poor finite size convergence
of the surface roughness inside the rough phase.

The scaling properties of the $p=q$ faceting transition
follow from the behaviour of the active bond density.
Numerically, the surface width scales at $p=q$ as $W \sim L$,
like in the faceted phase. From the perspective of the rough phase
the faceting transition takes place when the total active bond density,
$\bar\rho =\frac{1}{L} \int_0^L \rho(x) dx$, vanishes, because
in the faceted phase the $\xi_0$ segments of the rounded hill tops
are of measure zero compared to the ramp segments.
We find $\bar\rho \sim (q-p)^{\beta_\rho}$
with $\beta_\rho$ very close to 1.
At $p=q$ itself, the powerlaw $\rho(x) \sim x^{-1}$
predicts that $\rho$ scales with system size as $\bar\rho \sim \ln L / L$.
The numerical data in  Fig.~\ref{rho_xp=q}$(b)$ confirm this.
Finally, $\bar\rho$ decays at $p=q$ algebraically in time with exponent
$0.32(1)$.
This suggests a dynamic exponent $z= 3.1(2)$.

Erosion below the currently lowest exposed level becomes strictly forbidden at
$r=0$. The evaporation faceted phase becomes flat,  and a
directed Ising~(DI) type roughening transition takes place at
$p=p_{DI}=0.317(1)$.
Hinrichsen and Odor~\cite{HinOdo} already documented this.
They  independently introduced the $r=0$ limit of our model.
They also report that the surface width scales  
at $p_{DI}$ as $\sqrt{ \log(t)}$,
and inside the rough phase as $\log(t)$.

In summary, the presence of a
non-local topological constraint on equilibrium surface configurations
in dissociating dimer type surface growth,
leads to anomalous reduced surface roughness, with exponent
$\alpha=\frac{1}{3}$
instead of the conventional value $\alpha=\frac{1}{2}$.
Moreover, the growing (evaporating) surface is always faceted, due to the
spontaneous creation of pinning valleys (hill tops).
Under other circumstances, in particular when the
digging probability on flat surface segments is being suppressed,
an intermediate slowly growing rough phase appears
with complex scaling properties and strong corrections to scaling,
due to the presence of large internal length scales.
This research is supported by NSF grant DMR-9700430,
by the KOSEF through the SRC program of SNU-CTP,
and by the Korea Research Foundation (98-015-D00090).

\narrowtext

\end{multicols}
\end{document}